# Redox oscillations in 18650-type lithium-ion cell revealed by *in operando* Compton scattering imaging

Kosuke Suzuki,[1, a)] Shunta Suzuki,[1] Yuji Otsuka,[1] Naruki Tsuji,[2] Kirsi Jalkanen,[3] Jari Koskinen,[3] Kazushi Hoshi,[1] Ari-Pekka Honkanen,[4] Hasnain Hafiz,[5] Yoshiharu Sakurai,[2] Mika Kanninen,[3] Simo Huotari,[4] Arun Bansil,[6] Hiroshi Sakurai,[1] and Bernardo Barbiellini[7, 6]

[1)] *Faculty of Science and Technology, Gunma University, 1-5-1 Tenjin-cho, Kiryu, Gunma 376-8515, Japan*
[2)] *Japan Synchrotron Radiation Research Institute, SPring-8, 1-1-1 Kouto, Sayo, Hyogo 679-5198, Japan*
[3)] *Akkurate Oy, Kaarikatu 8b, 20760 Kaarina, Finland*
[4)] *Department of Physics, University of Helsinki, P.O. Box 64, FI-00014 Helsinki, Finland*
[5)] *Department of Mechanical Engineering, Carnegie Mellon University, Pittsburgh, PA 15213, USA*
[6)] *Department of Physics, Northeastern University, Boston, MA 02115, USA*
[7)] *Department of Physics, School of Engineering Science, LUT University, FI-53850 Lappeenranta, Finland*

## ABSTRACT

Compton scattering imaging using high-energy synchrotron x-rays allows the visualization of the spatio-temporal lithiation state in lithium-ion batteries probed *in-operando*. Here, we apply this imaging technique to the commercial 18650-type cylindrical lithium-ion battery. Our analysis of the lineshapes of the Compton scattering spectra taken from different electrode layers reveals the emergence of inhomogeneous lithiation patterns during the charge-discharge cycles. Moreover, these patterns exhibit oscillations in time where the dominant period corresponds to the time scale of the charging curve.

Demand for lithium-ion batteries (LIBs) is increasing globally because LIBs offer advantages of high energy-density, limited self-discharging, and longer life-cycles compared to other rechargeable batteries[1,2]. These advantages have made them attractive power source for portable electronic devices, electric vehicles, and key components of smart grids[3]. Although the need for the development of high-performance LIBs is clear, fundamental problems still remain in enhancing battery performance. In LIBs, as the lithium ions shuttle between the anode and the cathode, the charge-discharge process is driven by the reduction-oxidation (redox) reactions at the anodes and cathodes. The redox reaction generally occurs inhomogeneously because the electrodes consist of complex structures involving active materials, binders, and conductive additives. Inhomogeneous reactions have been reported, for instance, via *in operando* synchrotron x-ray diffraction[4,5] and hard x-ray-based tomography with x-ray absorption-near-edge-structure[6,7] studies. Inhomogeneous reactions lead to performance degradation of LIBs.

Although x-ray absorption (XAS)[8,9], x-ray emission (XES)[9] and resonant inelastic x-ray scattering (RIXS)[10] techniques have been used to probe the redox mechanism in electrode materials *in situ* and *in operando* conditions, these techniques often require the use of custom-made laboratory cells[11]. Neutron diffraction techniques have been deployed for non-destructive visualization of inhomogeneous reactions through the change of the lattice constants in commercial LIBs[12]. However, it is difficult to distinguish local regions within the cell given the limited spatial resolution of neutron beams. The Compton scattering technique is unique among the spectroscopies as a window on the bulk redox mechanisms and for non-destructive probing of the lithiation/delithiation reactions with high spatial resolution. By combining high-energy x-ray Compton scattering experiments with first-principles calculations, we have investigated the redox mechanism in various cathode materials such as $Li_xMn_2O_4$[13,14], $Li_xCoO_2$[15] and $Li_xFePO_4$[16]. Moreover, we have developed a non-destructive technique for lithium batteries, which we refer to as Compton scattering imaging[17]. Using this technique for the commercial CR2032 battery, the migration of the lithium ions to the cathode was visualized under discharge, and the structural changes due to the volume expansion of the electrodes were revealed[17]. In addition, we have introduced a method using the lineshape of the Compton scattering spectrum ($S$-parameter), which is sensitive to changes in lithium concentration[18]. An $S$-parameter based analysis has revealed the lithium concentration in the electrodes of a VL2020 coin-type battery during a cycle[19] and the dependency of the cycle rate with an inhomogeneous displacement of ions[20]. Such an analysis has also identified the difference between fresh and aged cylindrical 18650 LIB cell with micro-scale spatial resolution[21,22]. In this study, the Compton scattering imaging is applied to a commercial 18650 cylindrical cell to observe inhomoheneous redox reactions in various specific layers of the cathode and anode materials through high-spatial resolution measurements.

The commercial 18650 LIB (model MH1) used in this study was made by LG Chem, Ltd. The cell is composed by a graphite anode and a Ni-rich $Li_x$(Ni, Mn, Co)$O_2$ (NMC) cathode separated by a polymer film. The anode is coated on a 0.015 mm thick Cu collector foil while the cathode is coated on a 0.025 mm thick Al collector foil. Including the collector, the total thicknesses is 0.19 mm in the anode and 0.15 mm in the cathode. These electrode thicknesses were checked by teardown analysis of a pristine cell. The nominal capacity of this cell is 3200 mAh and the operation window is 4.2V-2.5V.

The Compton scattering imaging was performed at the high-energy inelastic scattering beamline, BL08W, of SPring-8 synchrotron facility in Japan. The experimental details are described by Suzuki *et al.*[21]. The energy of the incident x-rays is fixed at 115.65 keV. The incident and collimator slits sizes are 5 μm in height, 750 μm in width and 500 μm in diameter. The state of charge of the sample was controlled using a potentiostat/galvanostat HAG1232m (HOKUTO DENKO CORPORATION). The Compton scattering x-ray energy spectrum of each component in the cell is measured by shifting the sample position along the *z*-direction, which is the direction perpendicular to the incident x-rays. The measurement time at one point is 60 seconds. From the measured Compton scattering x-ray energy spectra, the *S*-parameter is obtained using the following equation

$$S = \frac{\int_{-d}^{d} J\left(p_z\right) dp_z}{\int_{-l}^{-d} J\left(p_z\right) dp_z + \int_{d}^{l} J\left(p_z\right) dp_z}, \tag{1}$$

where the Compton profile $J(p_z)$ reflects the lineshape of the Compton scattering x-ray energy spectrum[23]. The variable $p_z$ is the electron momentum along the scattering vector. The value $d$ = 1 atomic unit (a.u.) gives the momentum range of the lithium contribution[21]. The value $l$ = 5 a.u. sets the total momentum range of our measurement[21]. Our results shows that the width of the Compton profile $J(p_z)$ narrows with lithium concentration. Therefore, the *S*-parameter can be used to estimate the state of charge of cathode materials as shown by Suzuki *et al.*[19] Related *S*-parameter studies have been used to characterize battery materials using positron annihilation spectroscopy[24,25].

Figure 1 shows the *S*-parameter distribution with a charge-discharge cycle obtained by changing sample position along the *z*-direction. As shown in Fig. 1(a) (green box), the measurement region is from the outer steel to the anode of the second layer. This region corresponds to 0.75 mm in the *x*-direction and 0.42 mm in the *z*-direction. Fig. 1(b) (blue box), shows the inner structure of the measurement region obtained by Compton scattering x-ray intensity, which is defined as the number of counts/sec. Since Compton scattering x-ray intensity probes the electron density[26], the materials composing the cell are clearly delineated as illustrated in Fig. 1(b).

During the measurement, the cell is charged with a constant current (CC) mode of 350 mA until the 4.2 V cut-off voltage is reached. Then, the charging is continued in a constant voltage (CV) mode at 4.2 V for 3 hours. The discharge of the cell is performed with a CC of -350 mA until the 2.5 V cut-off voltage. Then, a CV mode at 2.5 V for 3 hours is continued. These voltage curves are shown in the upper panels of Figs. 1(c) and 1(d) highlighted by the magenta box. The *S*-parameter distribution as a function of the internal height and the charge and discharge time is shown in lower parts of Figs. 1(c) and 1(d). These distributions are obtained at intervals of 2.3 hours per line. Our observations show that the position of each component of the cell is shifted by about 30 μm along the *z*-direction by charging and discharging due to volume change induced by the lithium intercalation/deintercalation process.

Figure 2 shows the averaged *S*-parameter at the anode and the cathode with respect to charging and discharging. The *S*-parameter variation is about 1.3 % for the anode and about 2.1 % for the cathode. The upper axis in Fig. 2 provides the variation of lithium concentration in the graphite active material of the anode during the cycle. This lithium concentration is estimated by using two equations. The first equation gives the number of electrons $n_e$ which is equal to the lithium ions $n_{Li}$ displaced from the cathode, given by the formula

$$n_e = n_{Li} = \frac{I \cdot t}{F}, \tag{2}$$

where, $F = 9.6485 \times 10^4$ C/mol is the Faraday constant, $I$ is the constant current, and $t$ is the charging time, respectively, in the CC mode. Here, $I = 0.35$ A and $t = 8.67 \times 3600$ seconds. The second equation gives the number of carbon atoms $n_c$ in the graphite anode using the formula

$$n_C = \frac{V \cdot \rho}{W_C}, \tag{3}$$

where $V = 7.312$ cm$^3$ is an estimate of the volume of the anode, ρ is density of graphite (2.26 g/cm$^3$) and $W_c$ is the atomic weight of carbon (12.01 g/mol). Then, assuming $LiC_6$ is formed at the anode when the cell is fully charged, the corresponding lithium concentration is obtained by comparing the amount of lithium and carbon in moles. Since we do not know the percentage of inactive components in the graphite anode our calculation of the lithium concentration is not exact, but it provides a useful estimate.

In order to determine the lithiation state in each electrode on charging, the variation of the *S*-parameter has been examined by comparing two particular regions: one near the separator and another near the collector, as shown in Fig. 3(a). For the anode, the *S*-parameter monotonically increases with the exception of a small oscillation after 6.7 hours as shown in Fig. 3(b). For the cathode, the *S*-parameter near the collector monotonically decreases while the *S*-parameter near the separator presents large oscillation in time as shown in Figs. 3(c) and 3(d), respectively.

To better understand the nature of the large *S*-parameter oscillations, we focus on the upper cathode. Figure 4 shows the space-time *S*-parameter oscillation $\Delta S$ obtained by subtracting and interpolating from the *S*-parameter its average value in this region. One can notice that the redox reaction in this electrode produces remarkable features associated with $\Delta S$ and also with the lithium concentration, which are often referred to as Turing patterns after a pioneering paper by Turing[27], where he suggested that a system of chemical substances reacting together and diffusing through a material can form such patterns. Aragón *et al.*[28] have presented a model reaction-diffusion system with two species where the patterns oscillate in time as in the present case.

A Fourier analysis of $\Delta S$ given in Figure 5, allows us to visualize the scales in space and in time involved in the dynamics of $\Delta S$. Fig. 5(a) presents the wavelength distribution as a function of time while Fig. 5(b) shows the distribution averaged over time. Fig. 5(c) shows the period distribution as a function of vertical position whereas Fig. 5(d) illustrates the distribution averaged on vertical positions. The results contained in Figs. 4 and 5 indicate that the redox reactions when combined with diffusion produce a complex spatio-temporal behavior in the upper-cathode characterized by a distributions of frequencies and wavelengths. According to the power spectra, the dominating period is about 6 hours. This duration corresponds to the time scale of the charging curve and therefore it can be changed with the current. The dominating wavelength of this phenomenon is below 0.02 mm and it is related to the size of the grains of the active material. The reason for the appearance of these patterns is due to different mobility of lithium ions and electrons, and non-linear effects in the chemical reaction[28]. The *S*-parameter oscillation could be controlled by the charging and discharging speed of the LIBs. This implies that the cell can have an optimal cycle speed with a more homogeneous flow of ions.

In summary, this study applies the high-energy synchrotron x-ray Compton scattering imaging technique to a commercial 18650-type cylindrical LIB. We have visualized the spatio-temporal lithiation state of the electrode layers under *in operando* condition. The depth-resolved *S*-parameter analysis shows that the lithiation reaction produces Turing patterns oscillating in time. The dominating period corresponds to the time scale of the charging curve.

We thank Mr. Daisuke Hiramoto of Gunma University for his technical support of Compton scattering experiment. K.S. was supported by JSPS KAKENHI Grant Number JP19K05519 and partially supported by the Association for the Advancement of Science and Technology of Gunma University. Compton scattering experiments were performed with the approval of JASRI (Proposal Nos. 2019A1721 and 2019B1818). The work at Northeastern University was supported by the US Department of Energy (DOE), Office of Science, Basic Energy Sciences (grant number DE-FG02-07ER46352) and benefited from Northeastern University's Advanced Scientific Computation Center (ASCC) and the NERSC supercomputing center through DOE (grant number DE-AC02-05CH11231). A.-P.H. was supported by University of Helsinki Doctoral Program in Materials Research and Nanosciences (MATRENA). A.-P.H. and S.H. were supported by the Academy of Finland (grant no. 1295696).

## DATA AVAILABILITY

The data that support the findings of this study are available from the corresponding author upon reasonable request.

Figures

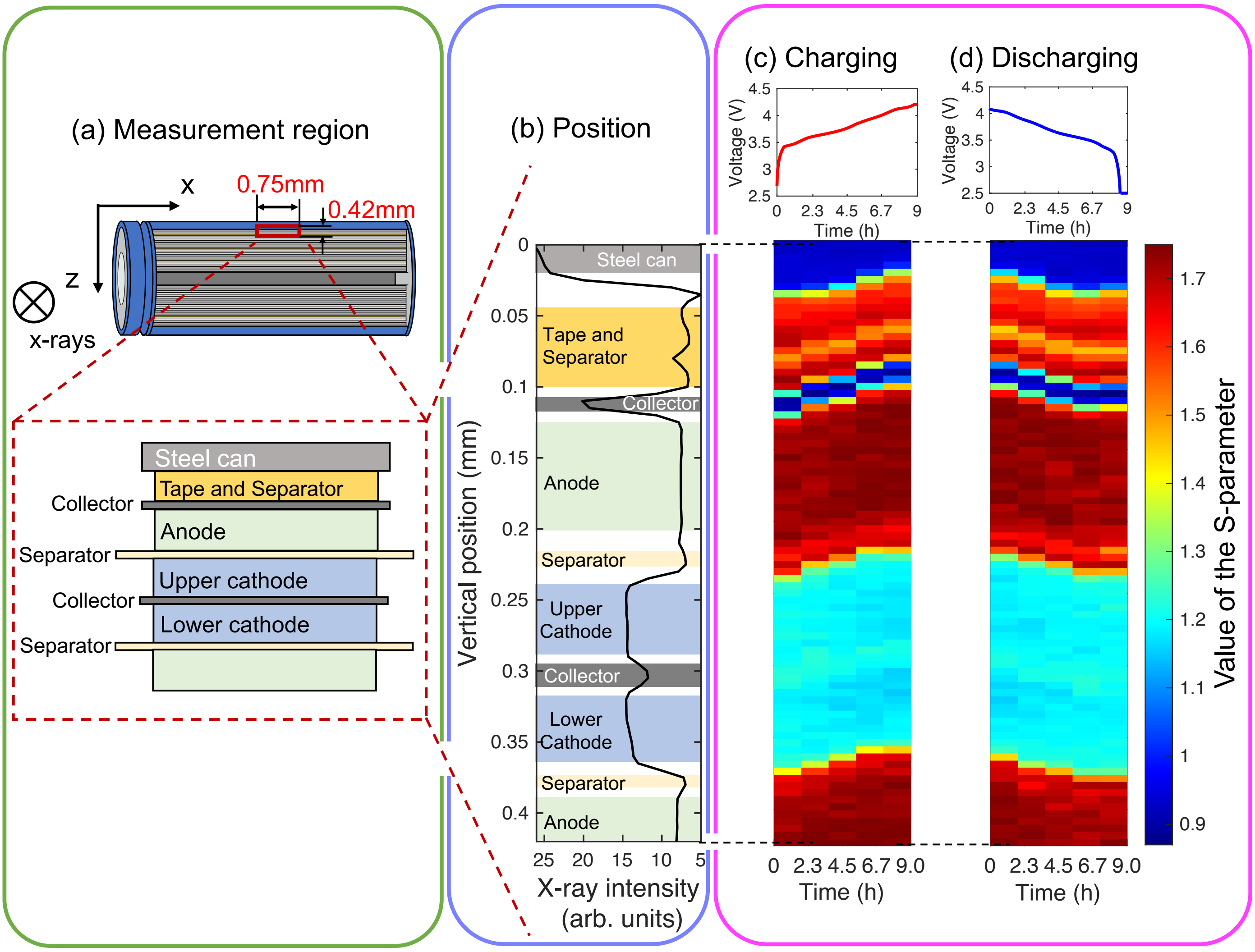

FIG. 1. Distribution of the *S*-parameter during a charge-discharge cycle. (a) schematic image of measurement region. (b) Inner structure at the measurement region obtained by Compton scattering x-ray intensity. (c) and (d) charge and discharge curves and the corresponding the distribution of the value of the *S*-parameter. The color of these figures indicates the value of the *S*-parameter.

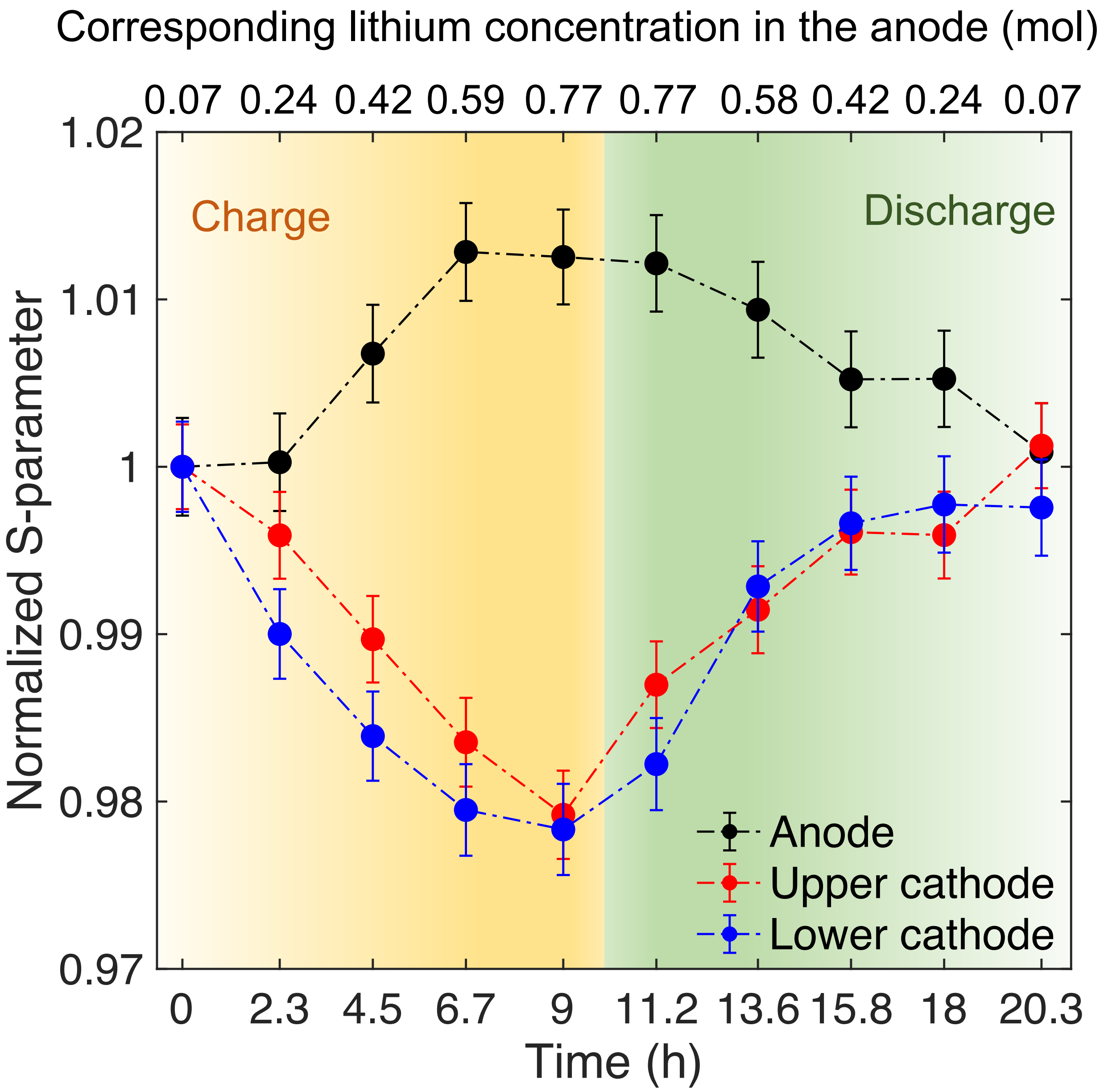

FIG. 2. Variation of the *S*-parameter at the electrodes during the charge-discharge cycle. The *S*-parameter s obtained from the anode, the upper cathode, and the lower cathode are shown in black, red, and blue circles, respectively. The background color corresponds to charging and discharging time. The cathode is composed of upper and lower parts as shown in Fig. 1(a). The time interval between the measured points is 2.3 hours.

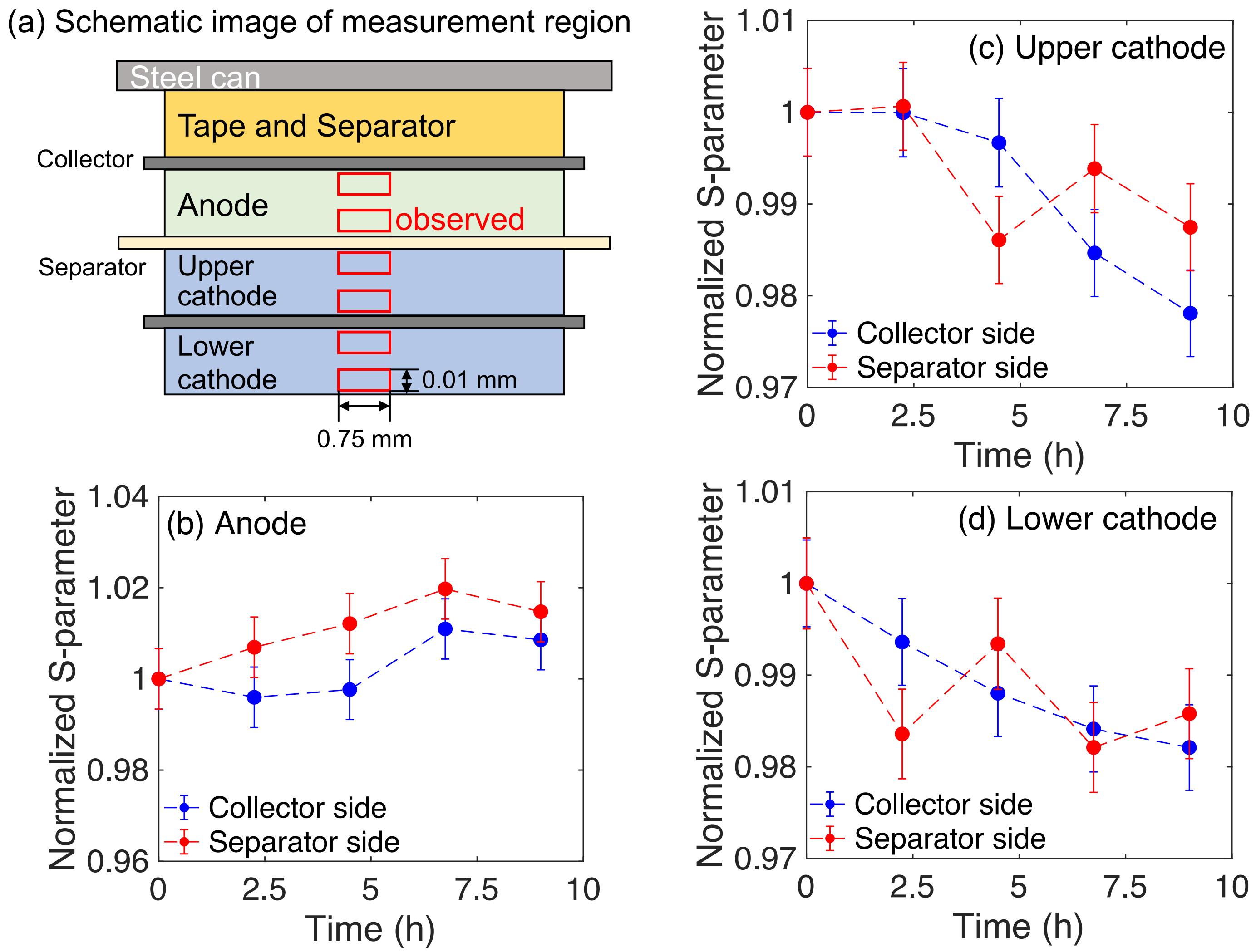

FIG. 3. Depth-resolved *S*-parameter analysis during the charge. (a) Schematic image of analysis region. (b) Normalized *S*-parameter obtained from the anode. (c) Normalized *S*-parameter obtained from the upper cathode. (d) Normalized *S*-parameter obtained from the lower cathode. The size of these regions delineated by red rectangles is 10 μm in height and 750 μm in width.

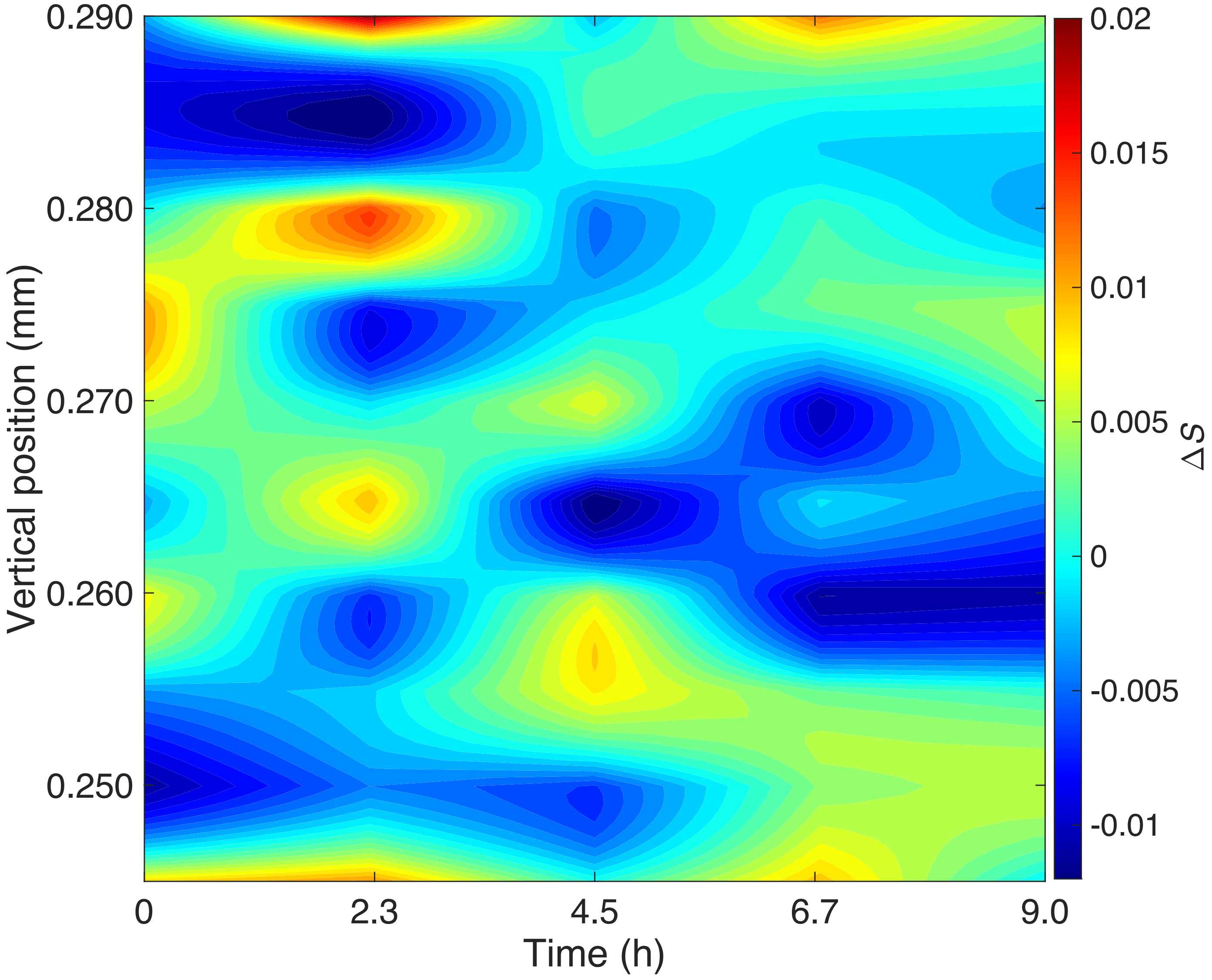

FIG. 4. Time and spatial oscillation component $\Delta S$ measured in the upper cathode during charging.

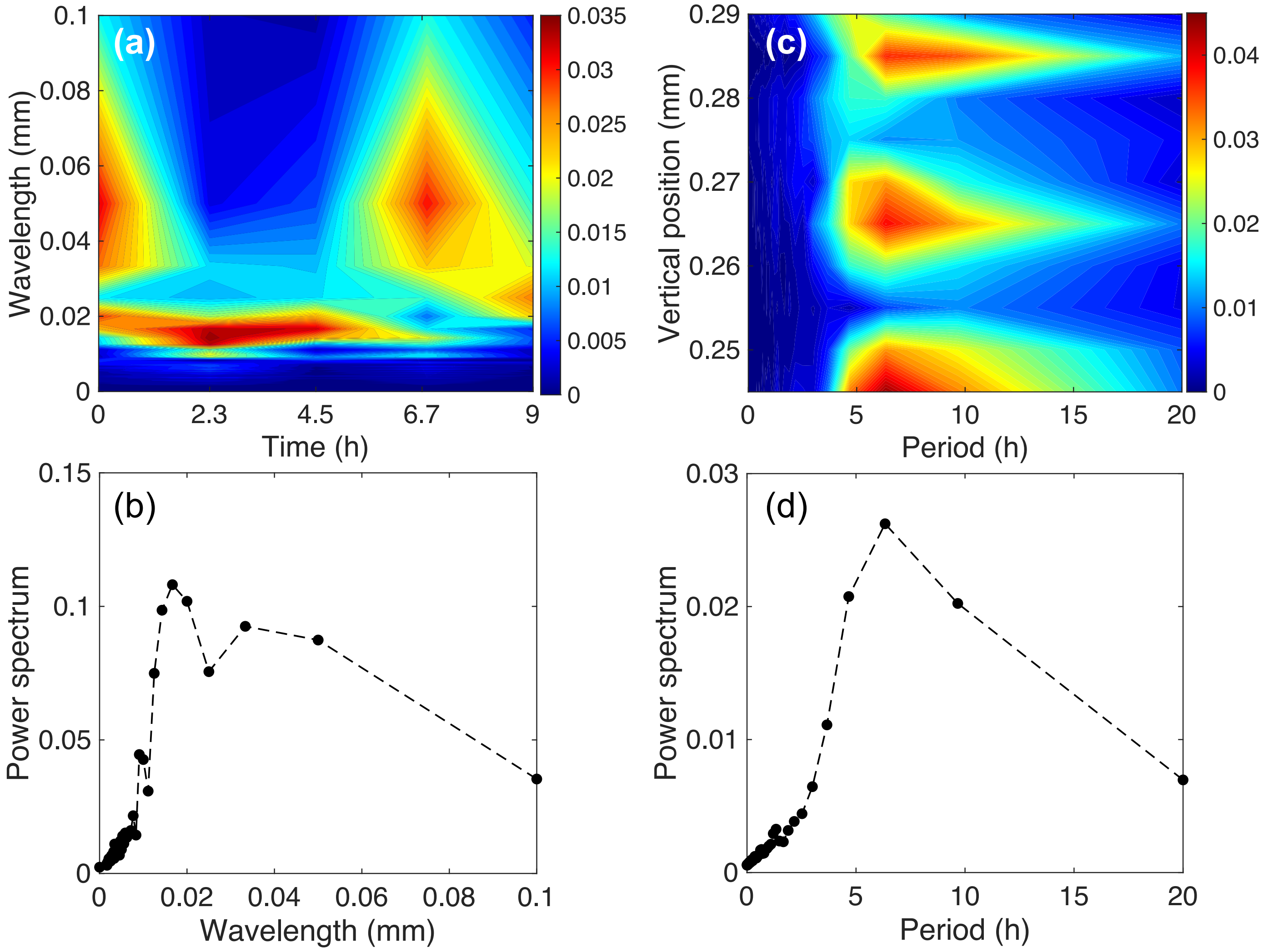

FIG. 5. (a) Time variation of the wavelength distribution of $\Delta S$. (b) Power spectrum for the wavelength distribution. (c) Spatial variation of the oscillation period distribution of $\Delta S$. (d) Power spectrum for the period.